\documentclass[aps,pre,twocolumn,float]{revtex4}

\usepackage{amsmath,bm,epsfig}

\newcommand{\Z}{\mathbb{Z}}

\newcommand{\ve}{{\varepsilon}}

\newcommand{\vp}{{\varphi}}

\newcommand{\pa}{{\partial}}

\DeclareMathOperator{\sgn}{sgn}

\usepackage[T1]{fontenc}

\usepackage[latin1]{inputenc}

\usepackage{amsmath}

\usepackage{amssymb}

\usepackage{amsfonts}

\usepackage{amsthm}

\def\a{\alpha}

\def\e{\varepsilon}

\def\o{\omega}

\let \nn  \nonumber

\def\be{\begin{equation}}       \def\ba{\begin{array}}

\def\ee{\end{equation}}         \def\ea{\end{array}}

\def\bea {\begin{eqnarray}}      \def\eea {\end{eqnarray}}

\def\bean{\begin{eqnarray*}}    \def\eean{\end{eqnarray*}}

\def\Fbox#1{\vskip1ex\hbox to 8.5cm{\hfil\fboxsep0.3cm\fbox{%
 \parbox{8.0cm}{#1}}\hfil}\vskip1ex\noindent}  
\let \nn  \nonumber

\def\RA {\ \Rightarrow\ }

\begin{document}

\title{Discrete wave turbulence of rotational capillary water waves}

\author{Adrian Constantin$^{\dag}$, Elena Kartashova$^\sharp$, Erik Wahl\'en$^{\ddag}$}
\affiliation{
$^\dag$ Faculty of Mathematics, University of Vienna, 1090 Vienna, Austria\\
$^\sharp$ RISC, J. Kepler University, Linz 4040, Austria\\
$^{\ddag}$ Centre for Mathematical Sciences, Lund University, PO
Box 118, 221\,00 Lund, Sweden;\\ FR 6.1 - Mathematik, Universit\"at des Saarlandes, Postfach 151150, 66041 Saarbr\"ucken, Germany}
\email{adrian.constantin@univie.ac.at, lena@risc.uni-linz.ac.at, erik.wahlen@math.lu.se}

\begin{abstract}
We study the discrete wave turbulent regime of capillary
water waves with constant non-zero vorticity. The explicit
Hamiltonian formulation and the corresponding coupling
coefficient are obtained. We also present the construction
and investigation of resonance clustering. Some physical
implications of the obtained results are discussed.
\end{abstract}

\maketitle

\section{Introduction}
Capillary water waves are free surface water waves, the range of wavelengths being
$
\lambda  < 17 \mbox{mm}.
$
In spite of their small amplitude, capillary waves play an important role in the dynamics
of the ocean surface. Wind generated capillary waves provide roughness on the ocean
surface which then yields subsequently the appearance of capillary-gravity and gravity
waves \cite{Li,PS,S}. On the other hand, capillary waves can be pumped by gravity waves \cite{NZ92,WB93}
and therefore they also transport energy from bigger to smaller scales.

The role of capillary water waves in the general theory of wave turbulence is invaluable. One can say
that most  new groundbreaking results and effects in (weak) wave turbulent systems have been
first discovered for capillary water waves. Seminal work of Zakharov and Filonenko \cite{zak2} laid in
1967 the very foundation for what later became known as statistical (or kinetic) wave turbulence theory:
the first power energy spectrum has been obtained for capillary water waves, as an exact solution of the wave kinetic equation.  The wave kinetic equation describes the wave field evolution at the finite inertial interval $0< (k_1,\, k_2) < \infty$ in $k$-space. At this interval the energy of a wavesystem is supposed to be constant while forcing (usually in the small scales $0<k_{for}<k_1$) and dissipation (in the scales $k_2 < k_{dis}$) are balanced \cite{ZLF92}.

Wave kinetic equations are deduced for the \emph{continuous} $k$-spectrum in
Fourier space while in bounded or periodical
systems the wave spectrum is \emph{discrete} and described by resonance clustering \cite{K09b}.
An important fact first shown by Kartashova in \cite{PRL94} is that independent resonance clusters
might exist in the infinite spectral space and are nonlocal in $k$-space \cite{PHD2,PRL94}.  Moreover, it was shown numerically (for atmospheric planetary waves) that increasing the wave amplitudes till some limiting magnitude yields no energy spreading from a cluster to all the modes of the spectral domain
but rather goes along some "channels" formed by quasi-resonances, that is, resonances with small enough frequency broadening (Fig.2, \cite{PRL94}). Non-resonant modes with comparatively big frequency broadening do not change their energy at the appropriate time scales (Fig.3, \cite{PRL94}).

A similar effect for irrotational capillary waves has been discovered by
Pushkarev and Zakharov \cite{Pu99,PZ99} who investigated how the discreteness of the spectrum
of irrotational capillary water waves influences the statistical properties of a weakly
nonlinear wave field described by the kinetic equation. It was shown in a series of numerical simulations with the dynamical equations of motion (with energy pumping into small wavenumbers) that the
wave spectrum consists of $\sim 10^2$ excited harmonics which do not generate an energy cascade
toward high wavenumbers --- \emph{frozen} or fluxless modes.
"There is virtually no energy absorption associated with high-wavenumbers damping in this case" (\cite{PZ99}, p.107). The quasi-resonance channels of the energy flux have also been observed and it was shown that the size of (quasi)-resonance clusters grows with increasing of the resonance broadening (\cite{PZ99}, p.113). Two mechanisms for increasing the number of flux-generating modes have been identified: 1) growing the modes` amplitudes, and 2) enlarging the number of modes.

The difference between the results of \cite{PRL94} and \cite{Pu99,PZ99} is as follows. Irrotational
capillary waves do not possess exact resonances \cite{K98} which yields the appearance of only fluxless modes; on the other hand, resonances do exist among atmospheric planetary waves. Accordingly, in \cite{PRL94} two types of modes had been singled out --- frozen and  flux-generating. The specifics of the
discrete wave turbulent regime is due to the fact that its time evolution is governed by the
dynamical equations, not by the kinetic equation \cite{K09b}; accordingly, phases of single modes are coherent and random phase approximation (necessary for deducing the wave kinetic equation) does not hold.

The pioneering result of Zakharov \emph{et al.}  \cite{ZKPD} was the discovery of the
\emph{mesoscopic} wave turbulence regime which
is characterized by the \emph{coexistence} of discrete and kinetic regimes in numerical simulations modeling turbulence of gravity waves on the surface of a deep ideal incompressible fluid.
The mathematical justification of this co-existence was given in  \cite{K06-1} in the frame of the model of laminated wave turbulence.

A peculiar property of \emph{irrotational} capillary waves is that they do not have exact
resonances in periodic boxes and therefore their discrete regime is exhibited in the form of frozen turbulence, with distinct modes just keeping their initial energy. As it was shown quite recently, \emph{rotational} capillary waves do have exact resonances, for specific positive magnitudes of the vorticity \cite{CK09}, which means that their discrete regime will have a different dynamics. The simplest
physical system where non-zero vorticity is important
is  that of tidal flows, which can be
realistically modeled as two-dimensional flows with constant
non-zero vorticity, with the sign of the vorticity distinguishing
between ebb/tide \cite{DP}.

The main goal of our paper is to study the discrete wave turbulent regime of \emph{rotational} capillary water waves with constant nonzero vorticity. In Section \ref{s:class} we give a brief overview of different turbulent regimes for three-wave resonance systems and we present known results on irrotational capillary waves
(with zero vorticity). Detailed presentation of  discrete wave turbulent regime is given in Section \ref{s:DWT}. In Section \ref{s:ham} we provide some background material on water waves with
vorticity.  Resonance clustering is constructed and studied in Section \ref{s:clust}. A brief discussion concludes the paper.

\section{Wave turbulent regimes}\label{s:class}

To introduce the three generic wave turbulent regimes we need to distinguish between the notions of quasi-resonance and an approximate (non-resonant) interaction.
Let us call a parameter $\delta_{\o},$
\be \label{quasi}
|\o_1 \pm \o_2 \pm ... \pm \o_s|=\delta(\o) \ge 0,  \ee
\emph{resonance width}. Regarding the dispersion function $\o_j$ as a function of \emph{integer} variables which are indices of Fourier harmonics (corresponding to a finite-size box), it was shown by \cite{K07}  that the resonance width can not be arbitrary small. This means that if $\delta(\o) \neq 0,$ then $\delta(\o) \ge \mathcal{R}(\o)$ where $\mathcal{R}(\o)$ is some positive number. Accordingly,
exact resonances, quasi-resonances and approximate interactions can be introduced as follows:

(1) \emph{exact resonances} are solutions of  (\ref{quasi}) with
\be \label{res-ex} \delta(\o) =0;\ee

(2) \emph{quasi-resonances} are solutions of (\ref{quasi}) with
\be \label{res-quasi} 0<\delta(\o) < \mathcal{R}(\o);\ee

(3) \emph{approximate interactions} are solutions of (\ref{quasi}) with
\be \label{res-non}\mathcal{R}(\o) \le  \delta(\o) < \mathcal{R}_{max}(\o),\ee
where $\mathcal{R}_{max}(\o)$ defines the  distance to the  nearest resonance. The number $\mathcal{R}(\o)$ can easily be found numerically, for a given dispersion function, as in \cite{Tan2004}.

\subsection{The model of laminated wave turbulence}

According to (\ref{res-ex})--(\ref{res-non}), discrete and continuous layers in a wave turbulent system are defined as follows.

The discrete layer is formed by exact resonances and quasi-resonances governed by (\ref{res-ex}) and (\ref{res-quasi}), while the continuous layer is formed by approximate interactions given by (\ref{res-non}). Four main distinctions between the time evolution of these two layers have to be accounted for:

1. dependence or independence on the initial conditions,

2. locality or non-locality of interactions,

3. existence or nonexistence of the inertial interval,

4. random or coherent phases.

\begin{center}
\emph{The continuous layer}
\end{center}
This layer is described by the wave kinetic equations and its stationary solutions in the form of power energy spectra. The main necessary assumptions in kinetic wave turbulence theory are \emph{random phase} approximation,  \emph{locality} of interactions and  existence of a finite \emph{inertial interval}. In the kinetic regime the time evolution of the wavefield \emph{does not depend on the details of the initial conditions} and not even on the magnitude of the initial energy. Only the fact that the complete energy of the wave system is conserved in the inertial interval is important. Interactions are \emph{local}, that is, only wave vectors with lengths of order $k$ are allowed to interact. Moreover, interactions take place only at some \emph{finite interval}, and no information is given about time evolution of the modes with wavenumbers lying outside this interval.

\begin{center}
\emph{The discrete layer}
\end{center}
This layer is governed by the set of dynamical systems describing resonance clusters; its time evolution \emph{depends crucially on the initial conditions}. Modes' phases are coherent;
the linear combination of the phases corresponding to the choice of resonance conditions (called \emph{dynamical phase} of a resonant cluster) describes
phases' time evolution (see Section \ref{ss:phas}).
Resonant interactions
\emph{are not local}, that is, waves with substantially different $k$-scales may form resonances. For instance, gravity-capillary  waves with wave lengths of order of $k$ and of the order of $k^3$  can form a resonance; more examples are given in \cite{K10}. Also the existence or absence of inertial interval is not important: resonances can occur all over the $k$-spectrum. Moreover, in some wave systems waves with \emph{arbitrary wavelengths} can form a resonance, for instance rotational capillary waves \cite{CK09}. We will discuss
this aspect in detail in Section \ref{s:clust}.

\begin{center}
\emph{Main aspects}
\end{center}

The main features of the model of laminated turbulence can be summarized as follows.

\textbullet~Independent clusters of resonantly interacting modes can exist \emph{all over $k$-spectrum}.

\textbullet~The time evolution of the modes belonging to a cluster are not described by KZ-spectra, thus leaving the "\emph{gaps}" in the spectra; their energetic behavior is described by finite (sometimes quite big) dynamical systems. Depending on their form and/or the initial conditions, the energy flux within a cluster can be regular (integrable) or chaotic.

\textbullet~The  places of the "gaps" in the spectral space are defined by the integer solutions of resonance conditions (e.g. see (\ref{3-res}) below). They are completely determined by the geometry of the wave system.

\textbullet~The size of the "gaps" depends on a) the form of dispersion function, b) the number of modes forming minimal possible resonance, and c) sometimes --- though not always --- also on the size of the spectral domain.

\textbullet~The discrete layer can be observable all over $k$-space and is characterized by the \emph{energy exchange} among a finite number of resonant modes. The continuous layer can be observable within the inertial interval $(k_1, k_2)$ and is characterized by the \emph{energy transport} over the $k$-scales.

\begin{figure}[t]
\vspace*{2mm}
\begin{center}
\includegraphics[width=9cm,height=6.5cm]{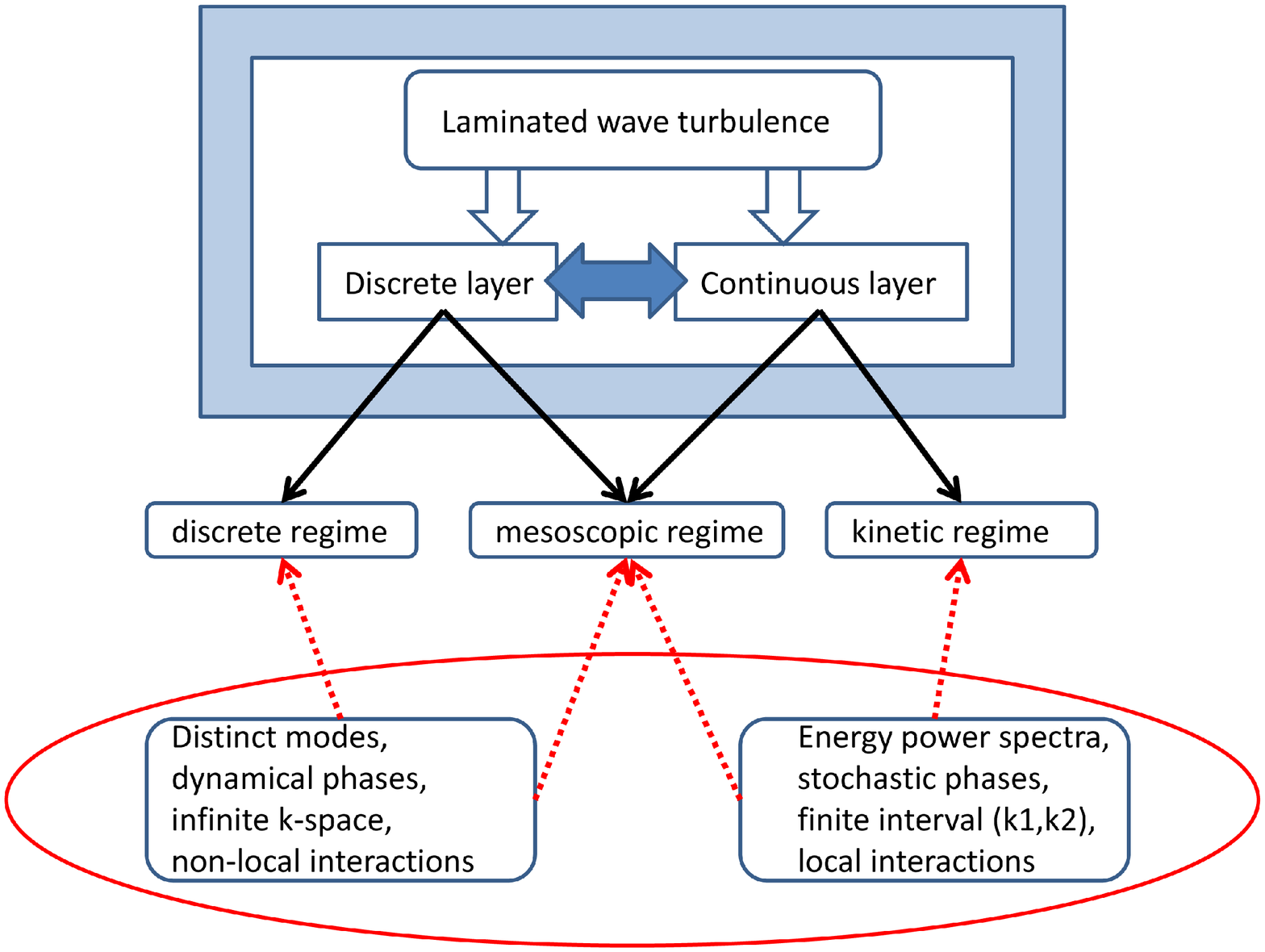}
\end{center}
\caption{\label{f:TurbTypes} Color online. Schematical representation of three wave turbulent regimes. In the upper panel two co-existing layers of turbulence are shown; (black) solid arrows are directed to the wave turbulent regimes in which a layer can be observable (shown in the middle panel).  (Red) dashed arrows going from the lower panel to the middle one show the evolutionary characteristics of the wavefield which manifest the corresponding regime.}
\end{figure}

Accordingly, \emph{three distinct wave turbulent regimes} can be identified with substantially different time evolution, as shown schematically in Fig.\ref{f:TurbTypes}. Energetic behavior of the modes in the \emph{discrete regime} is governed by a set of dynamical systems (see next section). 

Energy transport in the \emph{kinetic regime} is described by the \emph{kinetic} energy cascade, \be \label{kin-cas} E_{kin} \sim k^{-\nu},\ee where $\nu>0$ is a positive number depending on the dispersion function and the number of interacting waves, \cite{ZLF92}.

Much less is known about the \emph{mesoscopic regime} where both types of the wavefield time evolution can be detected \emph{simultaneously}: energy exchange within resonance clusters and energy cascades. The existence of energy-cascading clusters has been first shown in \cite{K10}. The simplest \emph{dynamic} cascade (without dissipation) has the form  \be \label{dyn-cas} E_{dyn} \sim p^{-n},\ee where $p$ is a constant, $0<p<1,$ $n=1,2,..., N$ and $N$ is the number of cascading modes within a cluster. Dissipation can also be included, i.e. dynamic energy cascade is not affected by the existence or non-existence of the inertial interval. 

Whether an energy cascade in the mesoscopic regime has kinetic or dynamic origin can easily be checked in experimental data.

In various physical systems either continuous or discrete or both layer(s) can be observable.
For instance, in the laboratory experiments reported in \cite{denis} only discrete dynamics has been identified. Coexistence of both types of the wavefield evolution has been detected in
laboratory experiments, \cite{WBP96}, and in numerical simulations, \cite{ZKPD}. Inclusion
of additional physical parameters could yield the transition from the kinetic to the
discrete or mesoscopic regime, \cite{CK09}.

\subsection{Parameter distinguishing among wave turbulent regimes}

In order to observe the various  wave turbulent regimes in laboratory experiments and in numerical simulations one has to carefully make a few choices.

First of all, it is necessary to define
the range of allowed amplitude magnitude so that the waves belong to the "corridor" of weak nonlinearity. This means that the amplitudes should be big enough to leave the linear regime but should not
be too big (to prevent strong turbulence). For water waves the wave steepness
$\e$, which is the characteristic ratio of the wave amplitude to the wavelength, is taken as a
suitable small parameter, the usually choice being $\e=0.1.$

Secondly, the choice of the spectral domain, i.e. the number of Fourier modes, is of the utmost importance. It
is usually regarded as the main parameter allowing to distinguish between different wave turbulent regimes. Namely, the ratio of the characteristic wavelength
$\lambda$ to the size of the experimental tank $L$ is, as a rule,
chosen as the appropriate parameter for estimating whether or not the waves "notice" the boundaries.
For capillary water waves, if $L/\lambda > 50,$ the wave system can be regarded as infinite and
the dispersion function, both in circular and in rectangular tanks, can be taken in the usual form as $k^{3/2}$ (M. Shats, private communication).
The different choices explain why various laboratory experiments with capillary water waves demonstrate different turbulent regimes. For instance, in \cite{SWW09} discrete modes have been identified  while in \cite{FFBF09} the development of full broadband spectra is observed.

\begin{center}
\emph{Characteristic wavelengths}
\end{center}
It is tempting to use the parameter
\be \label{rho} L/\lambda \ee
as a characteristic allowing to distinguish between different wave turbulent regimes: taking a sufficiently large
number of modes one will get the classical kinetic regime. However, the situation is more complicated. As it was shown in \cite{WBP96}, Fig.4,  low-frequency excitation (beginning with 25 Hz) generates distinct peaks of wave frequencies
while increasing the excitation to 300 Hz yields an isotropic turbulent regime with a
power law distribution of energies. Further increasing the excitation frequency generates an unexplained peak at 400 Hz. Moreover, capillary waves in Helium do notice the geometry of the experimental facilities for $L/\lambda \sim 30-300,$ the corresponding dispersion in circular tanks being described by Bessel functions (G. Kolmakov, private communication). In this case there exist nonlinear resonances and their manifestation has been recently established in laboratory experiments with  capillary waves on the surface of superfluid Helium, \cite{ABKL09}.
The local maximum of the wave amplitudes is detected at frequencies of the order of the
\emph{viscous cut-off frequency} in the case when the surface is driven by a periodical low-frequency force. This means that the energy is concentrated in a few discrete modes at the very end of the inertial interval $(k_1,k_2)$, in the beginning of the dissipative range.

These novel results are paramount both from theoretical and practical point of view. This is the first experimental confirmation of the predictions of the model of laminated wave turbulence that resonant modes can exist within and outside the inertial interval in the \emph{large} \emph{k}-scales.

 \begin{figure}[h]
\includegraphics[width=8cm,height=5cm]{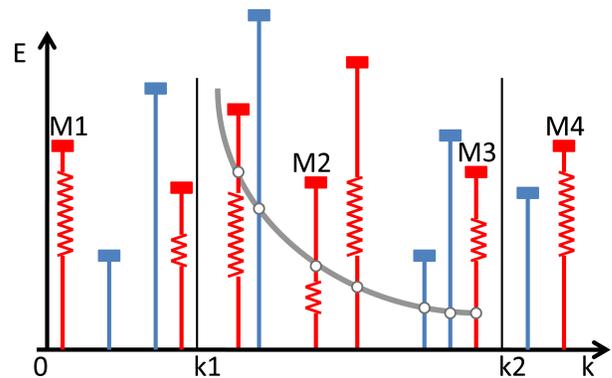}
\caption{\label{f:froz-vort} Color online.  The energy power spectrum $E_k$ is depicted  as the (grey)
   solid curve. Energies of resonant modes are shown by (red) vertical solid T-shaped lines having a string-like part.
   The energy of fluxless modes is shown by (blue) vertical solid T-shaped lines.
   The empty circles in
   corresponding
    places of the  energy power spectrum represent "gaps", indicating that for these modes the energy is
    given not by the solid curve but by the T-shaped lines. The horizontal and vertical axes show wavenumbers and energy correspondingly, while $k_1$ and $k_2$ denote the beginning and the end of the inertial interval.}
\end{figure}

The  origin of these peaks (similar to rogue waves in the oceans) in the wave spectrum  can be seen from the Fig.\ref{f:froz-vort}: whether or not a rogue wave is generated depends both on the driving frequency $\o_{drive}$ and on the resonant frequencies in a wave system. Indeed, suppose the modes $M_1, M_2, M_3$ have frequencies $\o_{1}, \o_2, \o_3$ and form a resonance $\o_{1}+ \o_2= \o_3$. If
$\o_{drive} \thickapprox \o_3$ (the mode with high frequency which is unstable), then the appearance of peaks with frequencies $\o_{1}$ and $\o_2$ is to be expected. If the modes $M_2, M_3, M_4$ form a resonance with $\o_2= \o_3+\o_4$, then an excitation with $\o_{drive} \thickapprox \o_2$ yields peaks at $\o_3$ and $\o_4$. The importance of the high-frequency mode is discussed in the next section.

Notice that rogue waves can appear within and outside the inertial interval while discrete wave turbulent regimes are not connected with the existence of the inertial interval. The situation is quite general and does not depend on the type of waves
under consideration. This is in agreement with the experimental observations in superfluid Helium where rogue waves have been detected and the fact that the energy balance is \emph{nonlocal} in $k$-space was established \cite{GEKMM08}.

\begin{center}
\emph{Characteristic resonance broadening}
\end{center}
The usual way to characterize the physical meaning of the resonance width $\delta_{\o}$ is
to regard it as a shift in the resonant wave frequencies, cf. \cite{PShX09,TB01,ZKPD}, yielding resonance broadening in a wide wave spectrum with a large number of modes.

To characterize the broadening in a three-wave system governed by (\ref{dyn3waves})
one can choose as an appropriate parameter
the inverse nonlinear oscillation time of resonant modes
\be
\tau_k^{-1} \approx |Z B_k|.
\ee
In discrete regimes the phases of individual modes are coherent; however if the width $|\Delta_{\o,B_3}|$ of the high-frequency mode in a resonant triad becomes substantially larger that
$\tau_k^{-1},$ the coherence is lost --- this being a necessary condition for the occurrence of the kinetic regime. One concludes then that

\noindent
\textbullet~The discrete regime corresponds to
\be \label{disc-regime} |\Delta_{\o,B_3}| \ll \tau_k^{-1},\ee
\textbullet~The kinetic regime corresponds to
\be \label{kin-regime}  |\Delta_{\o,B_3}| \gg \tau_k^{-1},\ee
\textbullet~The mesoscopic regime corresponds to
\be \label{meso-regime} |\Delta_{\o,B_3}| \simeq \tau_k^{-1}.\ee

In the case when exact resonances \emph{are absent} and only approximate interactions have to be accounted for,  the estimates (\ref{disc-regime})-(\ref{meso-regime})  can be made general (cumulative) by the substitution $|\Delta_{k}|$ instead of $|\Delta_{\o,B_3}|$, where $|\Delta_{\mathbf{k}}|$ is defined as a broadening of a mode with wave vector $\mathbf{k}.$
The fact that the generation of the kinetic regime occurs \emph{via} spectral broadening of discrete harmonics has been established in \cite{PShX09} for irrotational capillary waves.

There are three main reasons for the absence of exact resonances in a  wave turbulent system with decay type of dispersion function: a.  the resonance conditions do not have any solutions ($\o \sim k^{3/2}$); b.
the resonance conditions do not have solutions of a specific form (for example, for $\o \sim k^{-1}$
solutions exist in a square but not in a majority of rectangular domains); c. all resonances are formed in such a way that at least one mode in each primary cluster lies outside the inertial interval. In this case other modes will become "frozen" (\cite{Pu99,PZ99}) and no resonance occurs in the inertial range of the wavenumbers.
In the case when exact resonances are not absent, one has to be very careful while introducing any cumulative estimate:
for approximate interactions the upper boundary $\mathcal{R}_{max}(\o)$, which depends on $\mathbf{k}$, has to be included.
The problem of introducing a suitable cumulative parameter for distinguishing among various wave turbulent regimes
is contingent on future extensive research.

\section{Discrete regime}\label{s:DWT}
\begin{figure}[h]
\includegraphics[width=6.5cm,height=2.5cm]{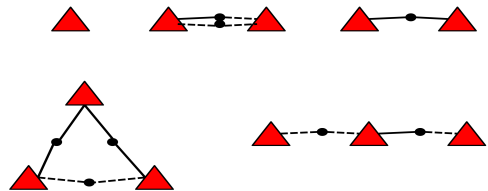}
\caption{\label{f:gaps} Color online. \emph{NR}-diagrams for clusters generating maximally 3, 4, 5, 6 and 7 gaps (from left to  right and from up to down).}
\end{figure}

The discrete wave turbulent regime is characterized by resonance clusters which can be formed by a big number of connected triads. The dynamical system corresponding to a cluster formed by $N$ triads
can be written out explicitly by  coupling the $N$ systems for a triad
\bea \label{dyn3waves}  \dot{B}_{1|j}=   Z_j B_{2|j}^*B_{3|j},
\dot{B}_{2|j}= Z_j B_{1|j}^* B_{3|j}, \nn \\ \dot{B}_{3|j}= -  Z_j B_{1|j} B_{2|j}, %
\eea
with $j=a,b,c,...$ and equaling the appropriate $B_{i|j}$. For instance, the dynamical system of the first two-triad cluster shown in Fig.\ref{f:gaps}  reads
\bea \label{kite-1}
\begin{cases} \nn
\dot{B}_{1|a}=   Z_a B_{2|a}^*B_{3|a},\,
\dot{B}_{2|a}= Z_a B_{1|a}^* B_{3|a}, \\ \dot{B}_{3|a}= -  Z_a B_{1|a} B_{2|a}, \,
 \dot{B}_{1|b}=   Z_b B_{2|b}^*B_{3|b},\\
\dot{B}_{2|b}=  Z_b B_{1|b}^* B_{3|b}, \, \dot{B}_{3|b}= -   Z_b B_{1|b} B_{2|b}, \quad \RA \\
B_{1|a}=B_{1|b} \,,  \, B_{3|a}=B_{2|b} \, %
\end{cases}
\eea
\vskip -0.5cm
\bea \label{kite-2}
\begin{cases}
\dot{B}_{1|a}=   Z_a B_{2|a}^*B_{3|a} + Z_b B_{3|a}^*B_{3|b},\\
\dot{B}_{2|a}= Z_a B_{1|a}^* B_{3|a}, \\
\dot{B}_{3|a}= -  Z_a B_{1|a} B_{2|a} + Z_b B_{1|a}^* B_{3|b},\\
\dot{B}_{3|b}= -   Z_b B_{1|a} B_{3|a},
\end{cases}
\eea
The system (\ref{kite-2}) is \emph{unique}, up to the change of indices $1 \leftrightarrow 2$. On the other hand, the form of the system changes depending on the fact whether or not the connecting mode is a high-frequency mode in one or both or
in no triads.

The high-frequency modes $B_{3|a}$ and $B_{3|b}$  are called \emph{active}, or $A$-modes, and the
other modes ($B_{1|a}, B_{2|a}, B_{1|b}, B_{2|b}$) are called \emph{passive}, or $P$-modes \cite{KL08}, because due to the Hasselman's criterion of instability \cite{Has67} $A$-modes are unstable under infinitesimal excitation and  $P$-modes are neutrally stable. The cluster dynamics is then defined by three possible connection types within the cluster, namely $AA-$, $AP-$ and $PP-$connections. In the system (\ref{kite-1}) the
connection $B_{1|a}=B_{1|b} $ is of $PP$-type and $B_{3|a}=B_{2|b}$ is of $AP$-type.

The graphical representation of an arbitrary resonance cluster in the form of a \emph{NR}-diagram (\emph{NR} for nonlinear resonance), first introduced in \cite{K09b}, allows us to keep this dynamical information. The vertices
of a \emph{NR}-diagram are triangles corresponding to resonant triads and the half-edges drawn as bold and dashed lines denote $A$- and $P$-modes correspondingly.
Examples of \emph{NR}-diagrams are shown in Fig. \ref{f:gaps}.

Each mode of a cluster generates a gap in the KZ-spectrum. The \emph{maximal possible number} of gaps can also be seen from the form of the $NR$-diagram, though the actual number can be smaller, in the case when two different wavevectors $k_1$ and $k_2$ have the same length.

An illustrative example can be found in \cite{CHS96} where laboratory experiments with two-dimensional gravity-capillary waves are described. By analyzing the mode frequencies in the measured data, only \emph{five} different frequencies were identified: 10, 15, 25, 35 and 60 Hz. However, theoretical consideration allowed to conclude that in fact \emph{seven} distinct modes take part in the nonlinear resonant interactions, and the corresponding resonance cluster has the form of a chain formed by three connected triads. The amplitudes and frequencies were identified as
\bea A_1 \leftrightarrow 60 \mbox{Hz}, \ \ A_2 \leftrightarrow 35 \mbox{Hz}, \ \ A_3 \leftrightarrow 25 \mbox{Hz},\nn \\ \quad A_4 \leftrightarrow 25 \mbox{Hz}, \ \ A_5 \leftrightarrow 10 \mbox{Hz}, \ \ A_6 \leftrightarrow 25 \mbox{Hz}, \ \ A_7 \leftrightarrow 15 \mbox{Hz},\nn \eea
and resonance conditions for frequencies as
\be
\o_1=\o_2+\o_3, \quad \o_2=\o_4+\o_5, \quad \o_6=\o_5+\o_7
\ee
"The waves $A_3, \, A_4$ and $A_6$ all have the same frequency (25 Hz), but they must have different  wavevectors, in order to satisfy the kinematic resonance conditions. We \emph{assume} that the mechanical means to generate a test wave at 25 Hz also generates perturbative waves in other directions at 25 Hz" (\cite{CHS96}, p.70).
\begin{figure}[h]
\begin{center}
\includegraphics[width=2.8cm,height=0.4cm]{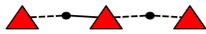}
\caption{\label{f:Diane} Color online. \emph{NR}-diagram of the resonance cluster appearing in the laboratory experiments with gravity-capillary water waves reported in \cite{CHS96}.}
\end{center}
\end{figure}
The \emph{NR}-diagram of this cluster is shown in Fig.\ref{f:Diane}: the cluster is a 3-chain with one \emph{PA}- and one \emph{PP}-connections. The dynamical system corresponding to the cluster has been solved numerically:
for all calculations the dynamical phase of
the initially excited triad was set to $\pi,$ while during the experiments the phases of the modes
have not been measured (D. Henderson, private communication). This yielded qualitative agreement of the results of numerical simulations with measured data but higher magnitudes of observed amplitudes.

The evolution of the amplitudes and phases in the discrete turbulent regime is discussed in Section \ref{ss:Amp} and
in Section \ref{ss:phas} correspondingly.

\subsection{Evolution of the amplitudes}\label{ss:Amp}
In Fig.\ref{f:froz-vort} we show the difference in dynamics of discrete modes of capillary waves with and without vorticity.
The energies of  fluxless modes in the frozen turbulence regime for irrotational capillary waves are shown as (blue) bold vertical lines and the energies of resonant modes appearing due to the non-zero constant vorticity are depicted by (red) bold lines with a spring part inside. The "springs" show symbolically that the magnitudes of energies can change in time, similarly to the process of compressing or pulling an elastic string. Notice that this situation is quite general and does not depend on the wave type (see \cite{PRL94} for numerical simulations of atmospheric planetary waves --- in Fig.1 the periodical
time evolution of resonant modes is shown and in Fig.3 the fluxless modes demonstrate no time changes in the magnitudes of amplitudes).

The analogy with a string is quite helpful for the understanding of the dynamics of resonant modes. A string can be compressed or pulled (without destruction) only in a finite range of the applied forces; similarly, the range of amplitudes' changing is also finite. Indeed, the simplest way to show it is just to look at the smallest possible resonant cluster --- a triad. The dynamical system for a resonant triad (\ref{dyn3waves}) has the Hamiltonian
\be \label{triad-ham} H_T = \operatorname{Im}(B_1 B_{2} B_{3}^*)\,,\ee
and
two Manley-Rowe constants of motion \cite{MR56} in the form
\be  \label{laws3waves}
I_{23}=|B_2 |^2 + |B_3|^2,  \  I_{13}= |B_1 |^2 + |B_3|^2,
\ee%
providing the integrability of (\ref{dyn3waves}) in terms of the Jacobian elliptic functions
$\mathbf{sn}, \mathbf{cn}$ and $\mathbf{dn}$ \cite{book-triad}. Notice that (\ref{dyn3waves}) has 6 real variables (real and imaginary parts of the amplitudes $B_j, \ j=1,2,3$) and only 3 conservation laws (\ref{triad-ham}), (\ref{laws3waves}) which, generally speaking, are not enough for integrability. However, if we rewrite it
in the standard amplitude-phase representation  $B_j= C_j\exp(i \theta_j)$ as
\bea \label{Triad-A-Ph}
\begin{cases}
\dot{C}_1=Z C_2C_3 \cos \varphi,\\
\dot{C}_2=Z C_1C_3 \cos \varphi,\\
\dot{C}_3=- Z C_1C_2 \cos \varphi,\\
\dot{\varphi}= -Z\,H_T (C_1^{-2}+C_2^{-2}-C_3^{-2}),
\end{cases}
\eea
it   becomes clear immediately that in fact we only have 4 independent variables: three real amplitudes $C_j$ and one dynamical phase
\be \label{dyn-ph}\vp=\theta_1+\theta_2-\theta_3\ee
according to the choice of resonance conditions. In this section we will concentrate on the wave amplitudes while the importance of dynamical phases will be discussed in the Section \ref{ss:phas}. Making use of the addition theorems
$ \mathbf{sn}^2+ \mathbf{cn}^2=1$ and  $\mathbf{dn}+\mu^2 \mathbf{sn} =1$
(here $\mu$ is a known function of conservation laws, $\mu=\mu (H,I_{23},I_{13}$) and
of the amplitude-phase representation $B_j= C_j\exp(i \theta_j)$, one can easily get explicit expressions for the squared amplitudes:
\be \label{C1-complex}
C_j^2(t)=\a_{1j} +
\a_{2j} \cdot \textbf{dn}^2 \Big(\a_{3j} \cdot t \cdot Z, \, \a_{4j}\Big)-\a_{5j}
\ee
with the coefficients $\a_{1j}, ... , \a_{5j}$ being explicit functions of $I_{13}, \ I_{23}$  and $H_T,$ for
$j=1,2,3,$ \cite{BK09_3}. The energy of each mode $E_j$ is proportional to the square of its amplitude, $E_j(t) \sim C_j^2(t) \sim \textbf{dn}^2,$ i.e. its changing range is finite and defined by the initial conditions as $0 < \textbf{dn}^2 (t) \le 1.$

There is no analytical expression for the amplitudes (energies) of a cluster formed by two or more triads. On the contrary, as it was shown in \cite{BK09_1,Ver68a,Ver68b}, the corresponding dynamical systems are integrable only in a few exceptional cases. For instance, clusters formed by $N$ triads all having one common mode and with all connections of $AA$- or $PP$-type are integrable for \emph{arbitrary initial conditions} if $Z_i/Z_j= 1/2, 1$ or $2$. Some clusters are known to be integrable for \emph{arbitrary coupling coefficients} but only for some specific initial conditions.
On the other hand, already the smallest possible cluster of two connected triads can
present chaotic behavior \cite{K10}, though the modes energies are still bounded as the corresponding dynamical systems are energy conserving.

In the case of a generic cluster, the time evolution of the amplitudes has to be studied numerically and depends crucially on the initial conditions.
In order to decrease the number of degrees of freedom one has to use the fact that each triad has $3$ integrals of motion given by (\ref{triad-ham})(\ref{laws3waves}), i.e. $N$ \emph{isolated triads} have $2N$ conservation laws and $N$ Hamiltonians. A cluster formed by $N$ triads with $n$ connections has $(2N-n)$ conservation laws and one Hamiltonian. The integrability of a generic cluster depends on the connection types within a cluster and on the ratios of coupling coefficients $Z_j$. Some results of numerical simulations with two-triad clusters (fixed connection type, various coupling coefficients) can be found in \cite{BK09_1}.

\subsection{Evolution of the dynamical phases}\label{ss:phas}

As one can see from (\ref{Triad-A-Ph}), the time evolution of a resonant triad does
not depend on the individual phases of the resonant modes but on the dynamical phase $\vp$
corresponding to the resonance conditions. In this case, the closed expression for the evolution of dynamical phase,
first found in \cite{BK09_3}, reads
\be \label{eq: phase} \varphi(t) = \textrm{sign}(\varphi_0)\,\mathrm{\mathbf{arccot}}\Big(k_1\,
\mathrm{\mathcal{F}} \Big(\,k_2\,(t-t_0),\mu\Big)\Big)\,,\ee
where
$$k_1 = -\,\frac{\mu}{|H_T|} \left(\frac{2 K(\mu)}{Z\,\tau}\right)^3,\quad
k_2 = 2\,\frac{K(\mu)}{\tau}$$
and
$$\mathrm{\mathcal{F}}(x,\mu) \equiv \mathrm{\mathbf{sn}}(x,\mu) \,\mathrm{\mathbf{cn}}(x,\mu) \,\mathrm{\mathbf{dn}}(x,\mu),$$
while $\varphi_0$ is the initial dynamical phase and the constants $\mu, K(\mu), \tau$ are known explicitly
as functions of $I_{13}, \ I_{23}$  and $H_T.$

The time evolution of a generic cluster again does not depend on the individual phases but on the dynamical phases; the
corresponding equations for the dynamical phases have a form similar to the last equation of (\ref{Triad-A-Ph}).
\begin{figure}[h]
\begin{center}\includegraphics[width=2cm,height=0.5cm]{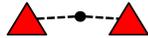}
\end{center}
\caption{\label{f:PP-but} \emph{NR}-diagram for a two-triad cluster with one \emph{PP}-connection.}
\end{figure}
For instance, in case of two triads
 $ a $ and $ b $ connected \emph{via} one \emph{PP}-connection (the \emph{NR}-diagram is shown in Fig.\ref{f:PP-but}),
 the equations for the dynamical phases
\be \label{ab-phases}
\varphi_{a} = \theta_{1|a} + \theta_{2|a} -\theta_{3|a}, \quad
\varphi_{b} = \theta_{1|b} + \theta_{2|b} -\theta_{3|b} \ee
read
 \bea \label{PP-standart}
\dot{\varphi}_a =  - H (C_{1}^{-2}+C_{2|a}^{-2}-C_{3|a}^{-2}),\\
\dot{\varphi}_b =  - H
(C_{1}^{-2}+C_{2|b}^{-2}-C_{3|b}^{-2}).
\eea
where the joint mode is denoted by $C_1$ and $H$ is the Hamiltonian
\be \label{m: Hamiltonian} H = C_{1} \left( Z_a C_{2|a} C_{3|a} \sin \varphi_a + Z_b
C_{2|b} C_{3|b} \sin \varphi_b\right).
\ee
\begin{figure}[h]
\includegraphics[width=4cm,height=5.5cm,angle=270]{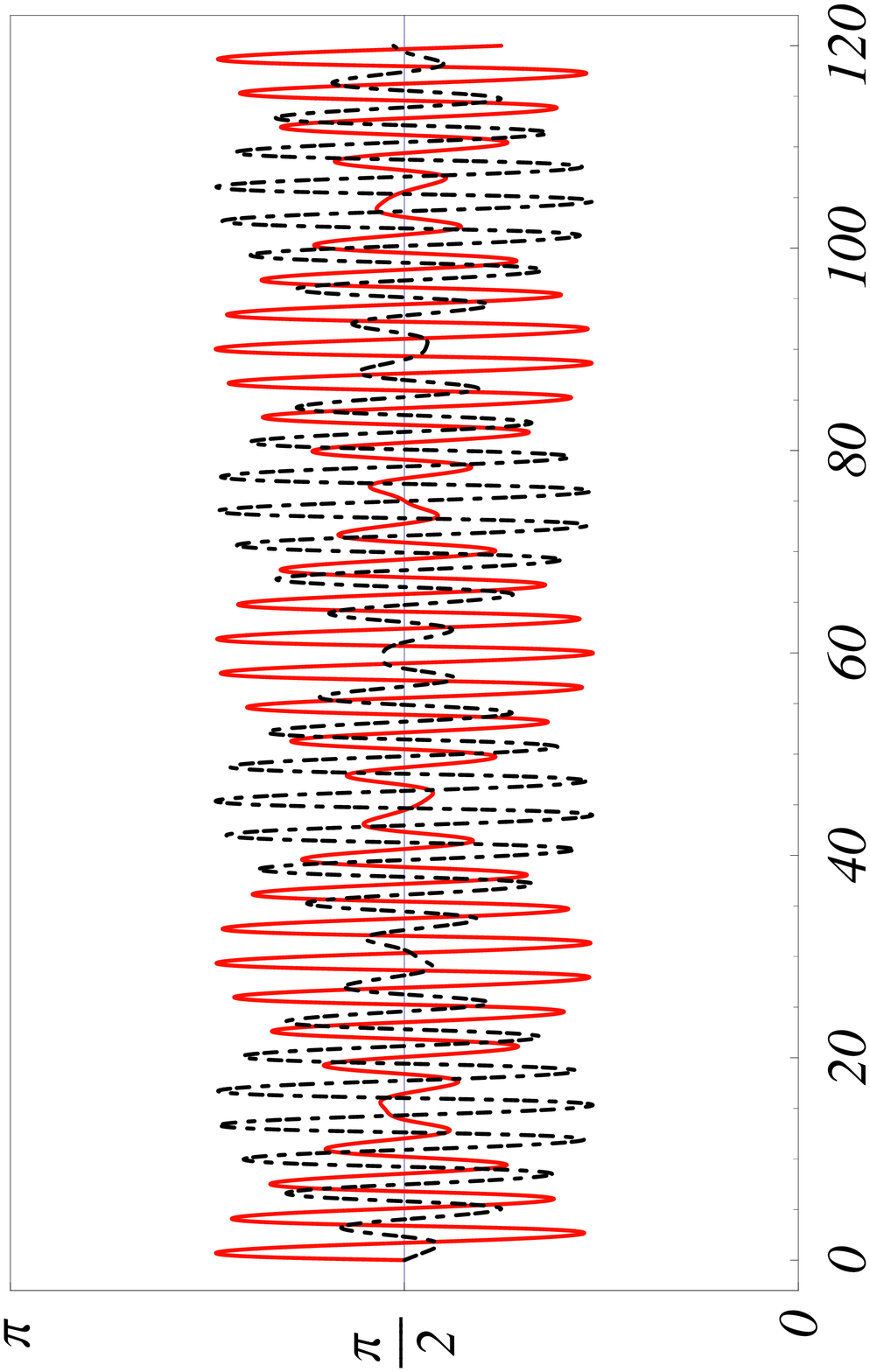}
\includegraphics[width=4cm,height=5.5cm,angle=270]{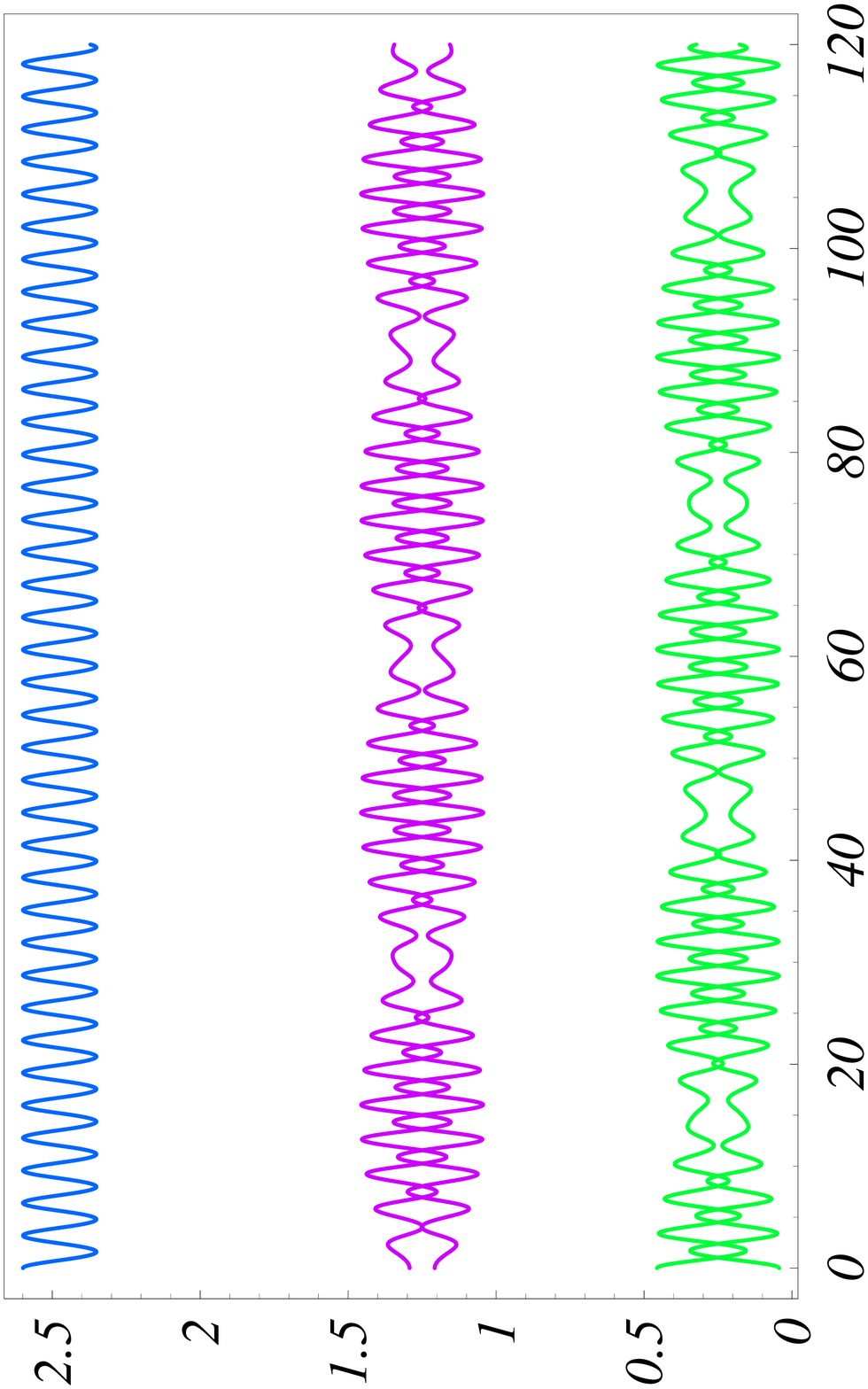}
\caption{\label{amp-ph-modul} \textbf{Upper panel}. Color online. Plots of the dynamical phases
as functions of time, for each
frame: $\varphi_a(t)$ is (red) solid and $\varphi_b(t)$ is (black)
dashed, with $\vp_{a,in}=\vp_{b,in}=\pi/2.$
\textbf{Lower panel}. Plots of  the squared amplitudes, as functions of time. The connecting mode $C_1$ is blue (the upper one),
$C_{2|a}$ and $C_{3|a}$ are green (in the middle), while $C_{2|b}$ and $C_{3|b}$ are purple (at the bottom).
To facilitate the view, $C_{2|b}^2,
C_{3|b}^2$ and  $C_1^2$ are shifted upwards. The horizontal axis stands for
non-dimensional time and the vertical axis for the phase in the upper panel and
for the squared amplitudes in the lower panel.}
\end{figure}
The effect of the dynamical phases on the evolution of the amplitudes
has been studied numerically in \cite{BK09_2}. It turned out that if initially $\vp_{in}=0$ for a triad, then it stays zero and the changing range of amplitudes is maximal. If $\vp_{in}$ is very small but non-zero, say, $\vp_{in}=0.01,$ then this range is substantially diminished: depending on the initial energy distribution within a triad it can decrease 10 times and more, being minimal for $\vp_{in}=\pi/2$. For the cluster shown in Fig.\ref{f:PP-but} the maximal range of the amplitudes is observed for $\vp_{a,in}=\vp_{b,in}=0$.

An interesting phenomenon has been uncovered in the case when at least one of the phases $\vp_{a}$ and $\vp_{b}$ is initially non-zero: a new time-scale is clearly observable where the amplitudes of the resonant modes are modulated by the evolution of the dynamical phases. The characteristic form of evolution is shown in Fig.\ref{amp-ph-modul} for $\vp_{a}=\vp_{b}=\pi/2,$ \cite{BK09_2}.

This is a manifestation of one more crucial distinction between  discrete and statistical wave turbulent regimes: while in the first regime phases are locked, in the latter they are supposed to be random. The effect of phase randomization among three-wave interactions of capillary waves has been studied recently experimentally in \cite{PShX09}.

The transition from a coherent-phase to a random-phase system occurs above some excitation threshold when coherent wave harmonics broaden spectrally. The waves were excited parametrically in the range of the shaker frequencies $40 \div 3500$ Hz and the modulation instability of capillary waves, which can be approximated by the squared secant function $\mbox{sech}^2$, has been established. Experiments show that the gradual development of the wave continuum occurs due to the spectral broadening. Similar effects have been previously observed in liquid Helium  \cite{RCS97} and in spin waves in ferrites \cite{KLM78}.

However, the form of the resulting KZ-spectrum does not allow to conclude automatically whether the energy cascade is due to three- or four-wave interactions, and both explanations can be found in the literature. A direct way to clarify the matter would be to estimate the periods of the excited modes for the case of three- and four-wave interactions at the corresponding time scale. The experimental confirmation can then be obtained by measuring the corresponding wave fields \cite{RP86}.

\section{Flows with constant vorticity}\label{s:ham}

A rotational water wave is a wave in which the underlying fluid flow exhibits non-zero vorticity. Physically rotational waves describe wave-current interactions: a uniform current is described by zero vorticity (irrotational flow) \cite{C, CE2, CS3}, while
a linearly sheared current has constant non-zero vorticity \cite{EV}. The irrotational flow setting is appropriate for waves
generated by a distant storm and entering a region of water in uniform flow \cite{Li}, while tidal flows are modelled by
constant vorticity \cite{DP}. The existence
of periodic gravity wave trains with small and large amplitudes \cite{CS, CS07} was established, and qualitative properties
of such waves were studied \cite{CEW07, CE1, CSS, CS2, CS07, GW, He, V08}. In the case of capillary- or gravity-capillary waves the rigorous theory is limited to the small-amplitude regime \cite{W07b}, although some partial results pertaining to the existence of large-amplitude waves have recently been obtained \cite{Wal09}. Note that already the small-amplitude theory is more involved in the presence of surface tension, including such phenomena as Wilton ripples \cite{W15} in which two harmonics with the ratio $1:2$ are in resonance.

An important new result on rotational capillary waves is the theorem on the dimension of flows with constant vorticity \cite{CK09}:\medskip

\textbf{Theorem}.
{\it Capillary wave trains can propagate at the free surface of a layer of water with a
flat bed in a flow of constant non-zero vorticity only if the flow is two-dimensional.}\medskip

This theorem allows us to consider for flows of constant vorticity only two-dimensional flows propagating in the $x$-direction.
For irrotational flows the existence of a velocity potential enables the transformation of the governing
equations for capillary water waves to a Hamiltonian system
expressed solely in terms of the free surface and of the
restriction of the velocity potential to the free surface \cite{Z99}.
The absence of a velocity potential for non-zero
vorticities complicates the analysis considerably and one can not expect results of this type.

However, for
flows of constant vorticity the remarkable preservation of the main features of the above theory (for irrotational
flows) can be established by introducing a generalized velocity potential. In particular there
is a Hamiltonian formulation in terms of two scalar variables, one of which is the free surface elevation \cite{W07}.
In contrast to the three-dimensional waves, it is
convenient  to regard the $y$-axis as the vertical axis, with the waves propagating in the
$x$-direction. The problem can be written in terms of a generalized velocity potential, satisfying
\begin{equation}\label{eq:phi}
\begin{cases}
\Delta \phi=0,& -\infty < y < \eta,\\
|\nabla \phi| \to 0, & y\to -\infty,\\
\end{cases}
\end{equation}
with $\phi_x=u+\Omega y$ and $\phi_y=v$, \cite{W07}. In this coordinate system
the free surface oscillates about the mean water level $y=0$.

Introducing the new variable $\zeta$, with
\be \zeta=\xi-\frac{\Omega}2\pa_x^{-1}\eta,\ee
where
$\xi=\phi|_{y=\eta}$ is the restriction
of the generalized velocity potential to the free surface, allows to
 obtain a canonical Hamiltonian system in  the form \cite{W07}:
\be
\dot \eta=\frac{\delta H}{\delta \xi},\quad
\dot \xi=-\frac{\delta H}{\delta \eta},
\ee
where the notation $\delta F/\delta u$ is used for the variational derivative
of a functional $F$ with respect to the variable $u$.

Passing to Fourier variables can be seen as a change of variables and
this transforms Hamilton's equations
into
\be
\dot \eta_k=\frac{\delta H}{\delta \xi_k^*},\quad
\dot \xi_k=-\frac{\delta H}{\delta \eta_k^*},
\ee
for $k\in \Z^*$.

The change of variables
\be
\eta_k=\sqrt{\frac{|k|}{2\tilde \omega(k)}}(a_k+ a_{-k}^*), \ \ \
\zeta_k=-i\sqrt{\frac{\tilde \omega(k)}{2|k|}}(a_k- a_{-k}^*) \nn
\ee
with
\be \label{om-til} \tilde \omega(k)=\sqrt{\sigma |k|^3+\frac{\Omega^2}{4}},\ee
linearizes the quadratic Hamiltonian and transforms Hamilton's
equations into
\be
a_k+i\frac{\delta H}{\delta a_k^*}=0.
\ee
In the new variables we have that
\be
H_2=\sum_{k\in \Z^*} \omega(k) |a_k|^2,
\ee
where
\be
\omega(k)=-\frac{\Omega}{2}\sgn k+\sqrt{\sigma|k|^3+\frac{\Omega^2}{4}},
\ee
is the dispersion relation for two-dimensional $2\pi$-periodic
capillary waves with constant vorticity \cite{W07b}.

The non resonant terms
in the Hamiltonian  can be  eliminated up to any desired order by the normal form transformation, i.e.
 we
can replace the Hamiltonian function $H=H_2+H_3+R_4$ by $\hat
H=H_2+H_3^\text{res}+\hat R_4$. Here $H_3^\text{res}$ is the resonant part
of $H_3$,

\be
H_3^\text{res}=\frac12 \sum_{\substack{k_3=k_1+k_2\\ \omega(k_3)=\omega(k_1)+\omega(k_2)}}
S(k_1,k_2,k_3)(a_{k_1} a_{k_2}a_{k_3}^*+c.c.),
\ee
with $S(k_1,k_2,k_3)$ being known functions of wavenumbers $k_j$ first found in \cite{CKW10}.

Let us now restrict our attention to a resonant
triad with resonance conditions taken in the form
\be \label{3-res}
\omega(k_3)=\omega(k_1)+\omega(k_2), \quad k_3=k_1+k_2. \ee

If we approximate the Hamiltonian using the quadratic and cubic terms,
Hamiltonian equations take the form
\[
\begin{cases}
\dot a_1+i\omega_1 a_1=-i S(k_1,k_2,k_3)a_2^* a_3,\\
\dot a_2+i\omega_2 a_2=-i S(k_1,k_2,k_3)a_1^* a_3,\\
\dot a_3+i\omega_3 a_3=-i S(k_1,k_2,k_3)a_1 a_2.
\end{cases}
\]
Setting $a_j=i B_j e^{-i\omega_j t}$, these equations transform into the standard form
\be
\dot B_1=Z B_2^*B_3,\ \
\dot B_2=Z B_1^* B_3,\ \
\dot B_3=-ZB_1 B_2,
\ee
where  the coupling coefficient reads, \cite{CKW10},
\be \label{Z}
Z=\sqrt{\frac{k_1 k_2
k_3}{\pi\tilde\omega_1\tilde\omega_2\tilde\omega_3}}\left(-\frac{\tilde\omega_1\tilde\omega_2}{2}+\frac{\Omega}{4}\tilde
\omega_3\right).
\ee

Note that in this formula the variables $k_j$, $\omega_j$, $\tilde \omega_j$ and $\Omega$ are not independent, since they are related by the resonance conditions \eqref{3-res}.
All possible magnitudes of the constant non-zero vorticity generating resonances, i.e. solutions of (\ref{3-res}), can be directly  computed, \cite{CK09}, as
\bea\label{om} \Omega (k_1,k_2) = \nn \\
\sqrt{\frac{\sigma}{6}} \cdot \frac{k_1 k_2\,(9k_1^2+9k_2^2+14k_1k_2)(k_1+
k_2)^{-1/2}}{\sqrt{(6k_1^4+15k_1^3k_2 +22k_1^2k_2^2 +15 k_1k_2^3+6k_2^4)}}\eea
for arbitrary  $k_1, k_2, k_3$ satisfying (\ref{3-res}).

It follows from (\ref{om})  the magnitude of a positive vorticity triggering a resonance can not be too small. The
minimal magnitude of the constant non-zero vorticity which generates resonances is
\be
\label{ommin}
\Omega_{min}=2\sqrt{\frac{\sigma}{3}}. \ee
The main conclusion one can deduce from (\ref{om}) is that any two rotational capillary
waves \emph{with arbitrary wavelengths} can form a resonance for a suitable magnitude
of vorticity. In other words, any vorticity computed by (\ref{om}) will generate an isolated triad; its dynamics has been discussed in detail in Section \ref{s:DWT}.

However, in laboratory experiments the magnitude of vorticity can only be controlled within some non-zero $\ve$ error
corresponding to the available accuracy of the experiments. For capillary waves $\ve \sim 10^{-2} $ is easy to achieve
and more refitments allow to reach  $\ve \sim 10^{-4} $ in some cases. In Section \ref{s:clust} we construct the resonance clustering of rotational capillary waves for various approximate magnitudes of the vorticity.

\section{Resonance clustering}\label{s:clust}
To see how the resonance clustering depends on the available accuracy $\ve$, let us define the
$\ve$-vicinity of vorticity $\Omega$ as
\be
\ve =\Big|\frac{\Omega_{max} -\Omega_{min}}{\Omega_{max}}\Big|.
\ee
and let us construct examples of clustering for a few different choices of $\ve$.

Notice that since the vorticity depends linearly on $\sqrt{\sigma}$, the previously defined
$\ve$ does not depend on $\sigma$ and consequently the
resonance clustering constructed below will be the same for capillary waves in an arbitrary liquid.
However, even for the same liquid, say water, the magnitude of $\sigma$ depends on the temperature and atmospheric pressure while performing experiments.
All  numerical simulations discussed below have been performed in the spectral domain $k_1, k_2 \le 100.$

\textbullet~ $10^{-8} \le \ve \leq 10^{-5}.$

We began  our numerical computations with $\ve=10^{-8}$ which does not generate any other clusters but one isolated triad for each magnitude of the approximate vorticity $|\Omega \pm \ve|$. Increasing  $\ve$ from $10^{-8}$ to $10^{-5}$ still leaves us with isolated triads.

\textbullet~ $10^{-5} \le \ve \leq 10^{-4}.$

For this values of $\e$, four new clusters appear, each formed by two connected triads:
\bea \label{ex1}
\begin{cases}
(20, 94, 114), (24, 70, 94); \\
(17, 71, 88), (15, 88, 103); \\
(11, 83, 94), (12, 71, 83); \\
(10, 47, 57), (12, 35, 47).
\end{cases}
\eea
Each line in (\ref{ex1}) consists of two resonance triads and the
exact magnitudes of the generating vorticities are given by (\ref{om}), say $\Omega(20,94)$ and $\Omega(24,70)$. These
magnitudes do not coincide, $\Omega(20,94) \neq \Omega(24,70)$, however
\be
|\frac{\Omega(20,94)-\Omega(24,70)}{\Omega(20,94)}| < \ve.
\ee

\begin{figure}[h]
\includegraphics[width=2cm,height=0.6cm]{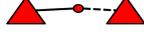}
\caption{\label{f:-4} NR-diagram for clusters appearing for $\ve \leq 10^{-4}.$ }
\end{figure}

All four two-triad clusters have the same NR-diagram shown in Fig.\ref{f:-4}.
No results on the integrability of corresponding dynamical systems are presently known, and
we expect chaotic energy exchange among the modes of such a cluster.

\textbullet~ $10^{-4} \le \ve \leq 10^{-3}.$

The structure of resonance clustering is substantially richer in this case. For instance, 83 different vorticities can generate two-cluster clusters, among those  most clusters have AP-connections but also clusters with AA-connections appear, for instance
\bea \label{ex2}
\begin{cases}
(50, 50, 100), (49, 51, 100); \\
(47, 48, 95), (46, 49, 95); \\
(44, 44, 88), (43, 45, 88)
\end{cases}
\eea
and others.
\begin{figure}[h]
\includegraphics[width=2cm,height=0.6cm]{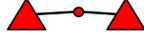}
\caption{\label{f:-3-2} NR-diagram for clusters described by (\ref{ex2}).}
\end{figure}

\noindent
The corresponding NR-diagram is shown in Fig.\ref{f:-3-2}. The
explicit form of the dynamical system for a cluster consisting of a
triad $a$ and a triad $b$ connected \emph{via} two high-frequency modes, $ B_{3|a}=B_{3|b}$,  reads
\bea \label{AA-but}
\begin{cases}
\dot{B}_{1|a}=  Z_{a} B_{2|a}^* B_{3|a}\,,\    \dot{B}_{1|b}=  Z_{b}  B_{2|b}^*B_{3|a}\,,\\
\dot{B}_{2|a}=  Z_a B_{1|a}^* B_{3|a}\,,\   \dot{B}_{2|b}=  Z_b B_{1|b}^* B_{3|a}\,,\\
\dot{B}_{3|a}= - Z_a B_{1|a}B_{2|a} -   Z_b B_{1|b}B_{2|b}.
\end{cases}
\eea
The integrability of (\ref{AA-but}) has been studied in \cite{Ver68a,BK09_1} with
the following results. The system (\ref{AA-but}) is integrable for \emph{arbitrary initial conditions}, if $Z_a/Z_b$ is equal to 1, 2 or 1/2, \cite{Ver68a}. In the case of arbitrary $Z_a/Z_b$, the integrability can be proven for some specific initial conditions \cite{BK09_3}.

Beside two-triad clusters we also observe the appearance of various structures formed by three, four, five and more connected triads; some examples  are given below.

\begin{figure}[h]
\includegraphics[width=2.5cm,height=1.5cm]{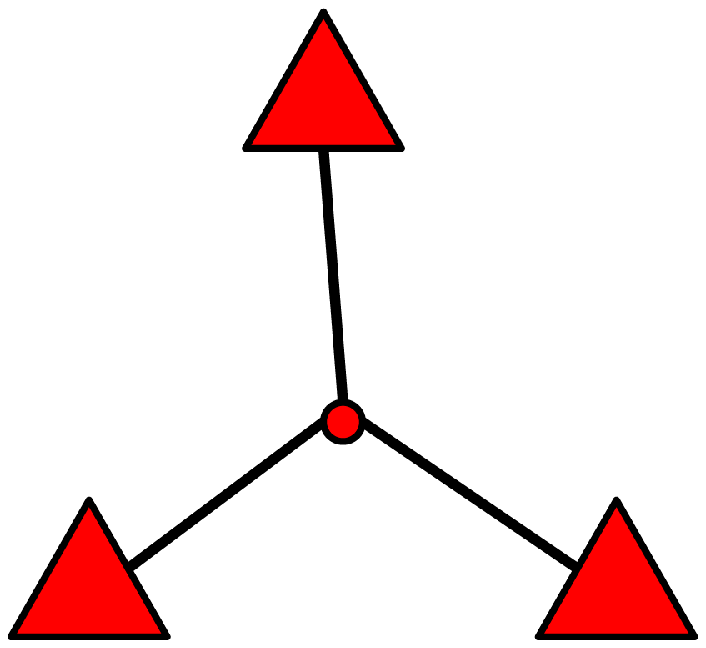}
\includegraphics[width=2.5cm,height=1.5cm]{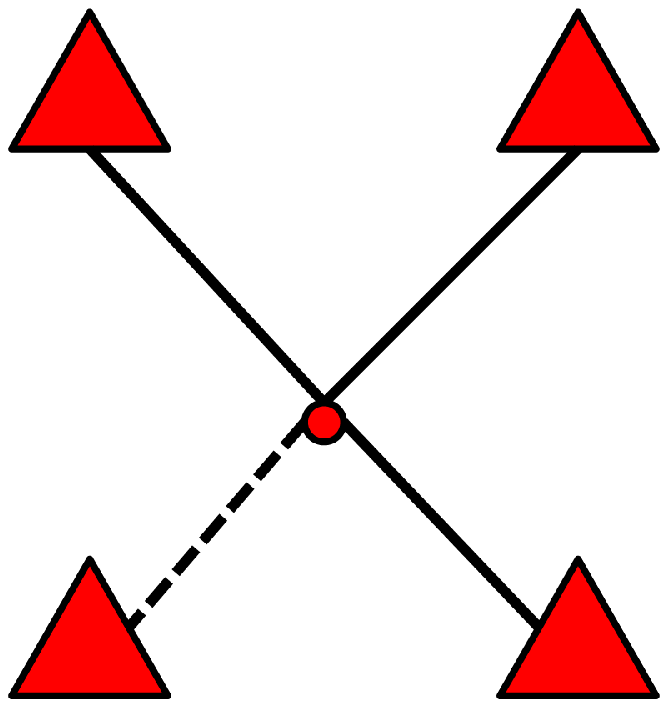}
\caption{\label{f:-3-34} NR-diagram for the clusters given by (\ref{3-star}) (on the left) and by (\ref{4-star}) (on the right).}
\end{figure}

The cluster  of three triads
\be \label{3-star}
(79, 80, 159), (78, 81, 159), (77, 82, 159)
\ee
has the NR-diagram shown in Fig.\ref{f:-3-34};
accordingly, its dynamical system reads
\bea
\begin{cases}
\dot B_{1|a}=Z_a B^*_{2|a}B_{3|a}\,, \ \dot B_{2|a}=Z_a B^*_{1|a}B_{3|a}\,, \\
\dot B_{1|b}=Z_b B^*_{2|b}B_{3|a}\,, \  ~ \dot B_{2|b}=Z_b B^*_{1|b}B_{3|a}\,, \\
\dot B_{1|c}=Z_c B^*_{2|c}B_{3|a}\,, \ ~\,  \dot B_{2|c}=Z_c B^*_{1|c}B_{3|a}\,,\\
\dot B_{3|a}=-Z_a B_{1|a}B_{2|a}   -Z_b B_{1|b}B_{2|b} -Z_c B_{1|c}B_{2|c}
\end{cases}
\eea
where the indices $a$, $b$ and $c$  are taken for three triads; all
connections within this cluster are of AA-type, i.e. $B_{3|a}=B_{3|b}=B_{3|c}$.

The cluster of four triads
\be \label{4-star}
(48, 48, 96), (47, 49, 96), (28, 96, 124), (46, 50, 96)
\ee
has three AA-connections and one AP-connection. The NR-diagram (see Fig.\ref{f:-3-34}) and
the dynamical system can be easily constructed but no results on the integrability of the corresponding dynamical system are known.

\textbullet~ $10^{-3} \le \ve \leq 10^{-2}.$

The size of the clusters grows exponentially with the growth of $\ve$ so that for $ \ve = 10^{-2}$ some vorticities generate clusters formed by a few hundreds to few thousands of connected triads. The
maximal cluster in the studied spectral domain consists of about 4000 triads with more than 33.000 connections among them.

It would be plausible to assume that in this situation no regular patterns will be observable for \emph{generic} initial conditions. On the other hand, a special choice of initial conditions --- excitation of a P-mode --- might produce a fluxless (frozen) regime. However,
one has to construct the resonance clustering and to check whether or not the "parasite" frequency of
electronic  equipment used during  experiments generates a resonance (see \cite{CHS96} for more detail).\\

Last but not least. Let us rewrite (\ref{om}) as
\be \label{om-rec1}
\Omega (k_1,k_2) =
\sigma^{1/2} \cdot \tilde{\Omega} (k_1,k_2)
\ee
where
\be\tilde{\Omega} (k_1,k_2) = \frac{k_1 k_2\,(9k_1^2+9k_2^2+14k_1k_2)(k_1+
k_2)^{-1/2}}{\sqrt{6(6k_1^4+15k_1^3k_2 +22k_1^2k_2^2 +15 k_1k_2^3+6k_2^4)}} \nonumber \ee
does not depend on the properties of fluid: these are absorbed in the coefficient $\sigma.$ The expression for $\tilde \omega(k)$ can also be rewritten as
\be \label{om-rec2} \tilde \omega(k)=\sqrt{\sigma |k|^3+\frac{\Omega^2}{4}}
= \sigma^{1/2}\sqrt{ |k|^3+\frac{\tilde{\Omega}^2}{4}},\ee
with $(|k|^3+\tilde{\Omega}^2/4)$ not depending on $\sigma.$ Correspondingly, the
interaction coefficient $Z$ has the form
\be \label{Z-rec}
Z=
\sigma^{1/4} \cdot \tilde{Z}
\ee
with $\tilde{Z}$ depending only on the wave numbers $k_1, k_2, k_3.$ This means that the cluster integrability defined by the ratios of the corresponding coupling coefficients \cite{Ver68a,Ver68b} also does not depend on the properties of fluid $\sigma$ while
\be
Z_{j_1}/Z_{j'}=\tilde{Z}_{j}/\tilde{Z}_{j'}
\ee
where $\tilde{Z}_{j}$ and $\tilde{Z}_{j'}$ do not depend on $\sigma$ and the indexes $j, j'$ correspond to
the $j$-th and the $j'$-th triad in a cluster. However, the magnitude of the vorticity $\Omega$ generating
the corresponding cluster will depend, of course, on the properties of a fluid as a function of $\sigma$.

Indeed,
$\sigma$ is ratio of the coefficient of surface tension $\tilde{\sigma}$ to the liquid density $\rho$, i.e. $\sigma=\tilde{\sigma}/\rho$,
and is different even for the same liquid if experiments are performed under different conditions
(such as a change of the ambient temperature). For instance, for water with
$\rho=998$  kg/m$^3$ and standard pressure,  $\tilde{\sigma}=72.14$ mN/m if the temperature of water is $25^{°}$ C and $\tilde{\sigma}=75.09$ mN/m if the temperature is $5^{°}$ C.  Accordingly, $\sigma=7.23 \cdot 10^{-5}$ and $\sigma=7.52 \cdot 10^{-5}$ for these two cases.

\section{Discussion}
Our main conclusions can be
formulated as follows:

\textbullet~Among the rotational capillary waves with constant non-zero vorticity resonances can occur only if 1) the flow is two-dimensional, and 2) the magnitude of vorticity is larger than $\Omega_{min}$ given by (\ref{ommin}).

\textbullet~Two arbitrary rotational one-dimensional capillary waves can form a resonance only
for appropriate magnitudes of the constant non-zero vorticity, with the exact magnitude given by (\ref{om}). Thus
a chosen magnitude of the vorticity generates one isolated resonance triad with a periodic energy exchange among the modes of
the triad and the magnitudes of the resulting amplitudes depend on the initial dynamical phase (\ref{eq: phase}). In a laboratory experiment where initially the A-mode of the triad is excited we expect the appearance of some regular patterns on the surface of the liquid.

\textbullet~If some non-zero experimental accuracy is taken into account, clusters of more complicated structure occur.
Their time evolution can be
regular (integrable) or chaotic, depending on the ratios of coupling coefficients and sometimes also on the initial conditions. For
an arbitrary cluster its dynamical system can be written out explicitly and solved numerically, thus predicting the results of a laboratory experiment beforehand.

\textbullet~The coupling coefficient for the corresponding dynamical system is given explicitly by (\ref{Z}).
Note that from of (\ref{Z}) it follows that in fact this expression gives the correct coupling coefficient for rotational capillary waves,
not only in water but in arbitrary liquid.

\textbullet~Last but not least. The discrete turbulent dynamics shown in Fig.\ref{f:froz-vort}
occurs for very general types of flow motions and is also valid for wave systems possessing 4-wave resonances though
the corresponding theoretical study is more involved. Due to Hasselman's criterion for 4-wave systems \cite{Has67}, there is no
analog to the $A$-mode in this case, while the excitation of \emph{any one mode} in a generic resonant quartet does not yield
an energy flow within a quartet. At least two resonant modes have to be excited. The only exception is given by the
quartets of the form $\o_1+\o_2=2\o_3$, which can be regarded essentially as a 3-wave resonance and
all the results presented above can be applied directly, with the $A$-mode having frequency $2\o_3$. The
dynamical system for a generic quartet with frequency resonance condition
\be
\label{res4}
\omega ({\bf k}_1) + \omega ({\bf k}_2)= \omega ({\bf k}_{3}) +\omega ({\bf k}_{4}),\\
\ee %
is also integrable in terms of Jacobian elliptic functions \cite{SS05}. The time
evolution of a quartet is defined by the fact whether or not two initially excited modes belong to the same side of the (\ref{res4}) or to the different sides. Correspondingly, the pairs $(\o_1,\o_2)$ and  $(\o_3,\o_4)$ are called 1-pairs, while the
pairs $(\o_1,\o_3),$ $(\o_1,\o_4),$ $(\o_2,\o_3),$ $(\o_2,\o_4)$ are called 2-pairs. The excitation of a 2-pair
again does not generate an energy exchange among the resonant modes, while the energy pumping into a 1-pair will produce it under some conditions (see \cite{K10}, Ch. 4, for more details). The $NR$-diagrams keep track of this dynamical information in the form of the edges \cite{K09b}.  Applying these results for gravity water waves (work in progress) we can see that in this case the
generation of rogue waves demands more conditions to be fulfilled than in a 3-wave system.


\begin{acknowledgements}
We acknowledge L.V. Abdurakhimov, D. Henderson, G.V. Kolmakov, S. Nazarenko and M. Shats for fruitful discussions. A.C. was supported by the Vienna Science and Technology Fund (WWTF). E.K. acknowledges
the support of the Austrian Science Foundation (FWF) under the project
P20164-N18 ``Discrete resonances in nonlinear wave systems". E.W. was supported by an Alexander von Humboldt Research Fellowship.
\end{acknowledgements}

\end{document}